\documentclass[twocolumn,showpacs]{revtex4}
\usepackage{dcolumn}
\usepackage{epsfig,epsf}
\newcommand{\la}{\langle}
\newcommand{\ra}{\rangle}
\newcommand{\cO}{{\cal O}}

\newcommand{\bee}{\begin{equation}}
\newcommand{\ee}{\end{equation}}
\newcommand{\bea}{\begin{eqnarray}}
\newcommand{\eea}{\end{eqnarray}}
\newcommand{\R}{\rm I\kern-.2emR}
\newcommand{\C}{\rm \kern.25em\vrule height1.4ex
depth-.12ex width.06em\kern-.31em C}
\newcommand{\N}{{\rm I\kern-.16em N}}
\newcommand{\Z}{{\rm Z\kern-.35em Z}}

\begin{document}                                                                
MPP-2004-75\\
\bigskip\bigskip
\title{
Some new results on an old controversy: is perturbation theory the 
correct asymptotic expansion in nonabelian models?
}
\author{\it Miguel Aguado and Erhard Seiler}
\affiliation{\it Max-Planck-Institut f\"ur Physik
 (Werner-Heisenberg-Institut),
 F\"ohringer Ring 6, 80805 Munich, Germany
 }
\begin{abstract}  
Several years ago it was found that perturbation theory for
two-dimensional $O(N)$ models depends on boundary conditions even after
the infinite volume limit has been taken termwise, provided
$N>2$. There ensued a discussion whether the boundary conditions
introduced to show this phenomenon were somehow anomalous and there was a
class of `reasonable' boundary conditions not suffering from this
ambiguity. Here we present the results of some computations that may be
interpreted as giving some support for the correctness of perturbation 
theory with conventional boundary conditions; however the fundamental 
underlying question of the correctness of perturbation theory in these 
models and in particular the perturbative $\beta$ function remain 
challenging problems of mathematical physics. 
\end{abstract}
\pacs {11.15.Ha, 11.15Bt}
\maketitle
\vskip2mm
\section{Introduction}
Several years ago A. Patrascioiu and the second-named author \cite{ps} 
investigated the dependence of perturbation theory (PT) on boundary 
conditions (b.c.). Surprisingly they found that in two-dimensional 
nonlinear sigma-models with non-abelian symmetry the coefficients of the 
perturbative expansion depend on the boundary conditions (b.c.) used, even 
after the infinite volume limit is taken term by term. In particular it 
was found that the so-called superinstanton b.c. (s.i.b.c.) lead to results 
that differ from the standard ones (typically obtained using periodic 
b.c.). This is remarkable, because the full, nonperturbative thermodynamic 
limit should be independent of the b.c. used to approximate it.

A little later F. Niedermayer, M. Niedermaier and P. Weisz \cite{nnw} on 
the one hand confirmed and sharpened the results of \cite{ps}: they gave 
an analytic expression for the second order term in the PT expansion of 
the energy and in addition showed that the third order term diverges in 
the thermodynamic limit, thereby in some sense exhibiting an even stronger 
dependence on the b.c. used. On the other hand they gave arguments why the 
s.i.b.c. are in some sense anomalous; using plausible but unproven 
correlation inequalities they argued that the thermodynamic limit of 
the invariant two-point function should be sandwiched between the finite 
volume two-point function with free (Neumann) and 0-Dirichlet b.c., 
whereas the s.i.b.c. two-point function lies outside that interval. 

Assuming certain bounds on the remainder terms of the truncated PT 
expansion and also assuming the equality of the termwise thermodynamic 
limits with free and Dirichlet b.c. the authors of \cite{nnw} conclude 
that with those b.c. one obtains indeed a correct asymptotic expansion of 
the infinite volume two-point functions by taking the infinite volume 
limit term by term. 
  
Patrascioiu and the second named author \cite{psc} wrote a comment on 
the paper \cite{nnw} in which they stressed that the arguments put forward 
in that paper did not suffice to settle the issue of the correctness of 
conventional PT, since several unproven assumptions were made.

The present paper answers at least one of the question left over, by 
demonstrating that at least one of the assumptions made in \cite{nnw} is 
indeed correct.

\section{The problem}
We are dealing with the $O(N)$ vector models in $2D$; these models 
describe `spins' (unit vectors in ${\R}^N$) assigned to the vertices of a 
(finite) lattice $\Lambda\subset\Z^2$ and distributed with a Gibbs measure 
$\exp(-S_\Lambda)d\mu_\Lambda$ defined in terms of the action 
\bee
S_\Lambda\equiv \beta \sum _{\la xy\ra} (1-\vec s_x\cdot \vec s_y)
\ee
and the $O(N)$ invariant a priori measure $d\mu_\Lambda$. One is of course 
interested in the thermodynamic limit $\Lambda\to\Z^2$, in practice taken 
through square lattices of size $L$.

In a nutshell the problem is the following: consider the expectation value 
$\la \cO\ra_{L,bc}$  of an observable $\cO$ in a finite box of 
size $L$ with boundary conditions $bc$. Then PT can be derived using 
standard theorems on asymptotic expansions of integrals and yields
\bee
|\la \cO\ra_{L,bc}-\sum_{i=0}^k c_{i,bc}(L)\beta^{-i}|=R_k(\beta,L) 
\label{PTL}
\ee
with 
\bee
\lim_{\beta\to\infty} R_k(\beta,L) \beta^k=0\ .
\ee
The problem is what happens to this asymptotic expansion in the 
thermodynamic limit $L\to\infty$. It may happen, and has been shown in 
the case of the $O(N)$ models for $O(N)$ invariant observables and 
periodic b.c. \cite{eli,david}, that 
\bee
\lim_{\beta\to\infty}c_i(L)\equiv c_i(\infty)
\ee 
exists for all $i$, but the unsolved problem is whether a 
$R_k(\beta,\infty)$ 
with $\lim_{\beta\to\infty} R_k(\beta,\infty) \beta^k=0$ exists, such that 
the analogue of (\ref{PTL}) holds with $L$ replaced by $\infty$. A 
proof of this would require some uniform control of the $R_k(\beta,L)$, 
which so far has not been achieved, except for $N=2$ \cite{bric}.
 
\begin{figure}[htb]
\centerline{\epsfxsize=8.0cm\epsfbox{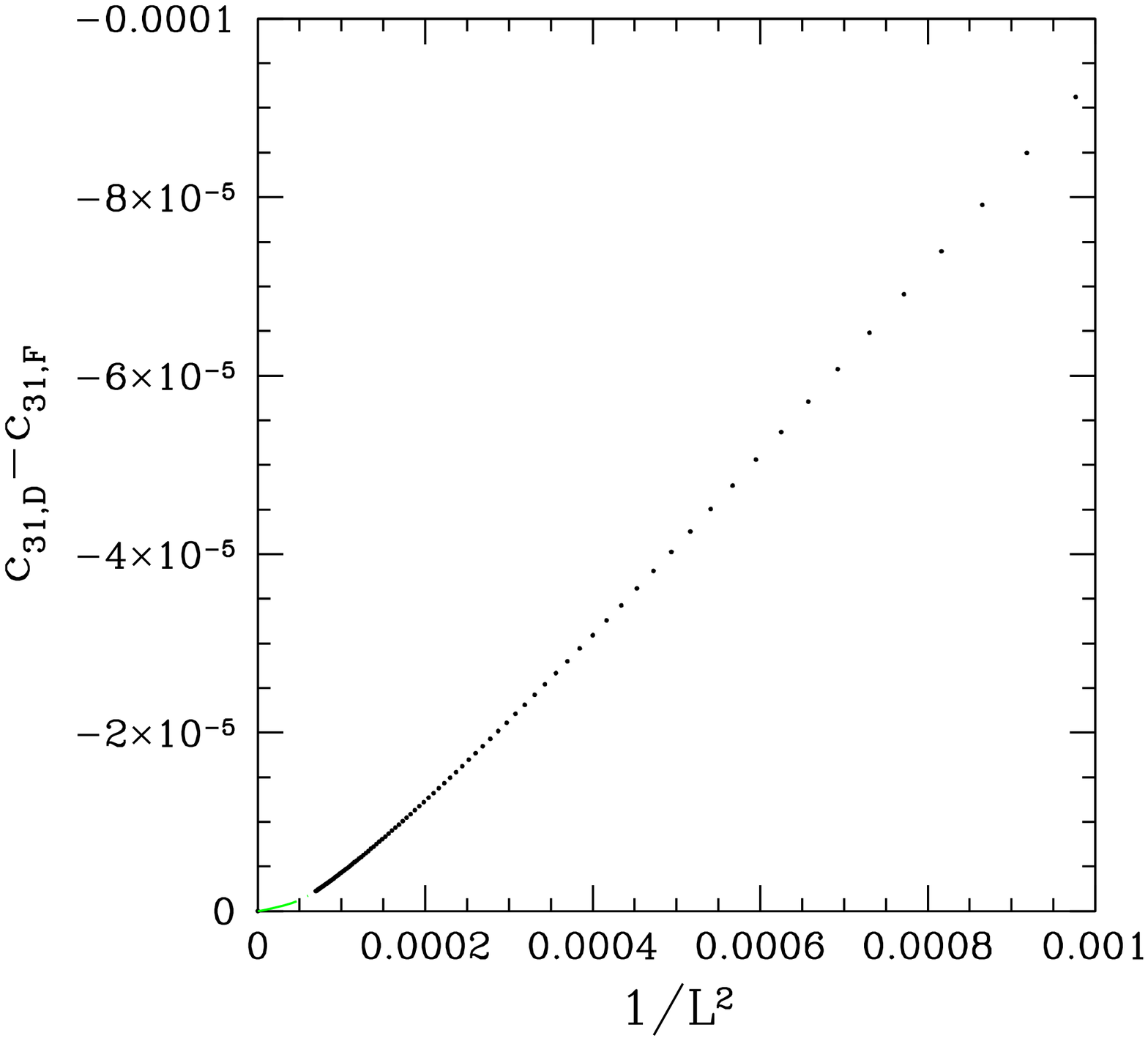}}
\caption{Coefficient $c_{31}$: difference between free and Dirichlet b.c.
}
\label{c31}
\end{figure}

\section{New results}
The idea of  \cite{nnw} is to sandwich infinite volume 
expectations between the finite volume ones with Dirichlet and free 
b.c..:
\bee
\la\cO\ra_{L,F}\leq \la\cO\ra_{\infty} \leq \la\cO\ra_{L,D}\ ,
\label{ineq}
\ee
for $\cO$ in a certain class containing invariant polynomials in the spins 
with positive coefficients (this includes in particular the invariant 
two-point functions). For $N=2$ (\ref{ineq}) is a consequence of 
Ginibre's inequalities \cite{gin}, whereas for $N>2$ is remains a 
plausible, but unproven conjecture. 

\begin{figure}[htb]
\centerline{\epsfxsize=8.0cm\epsfbox{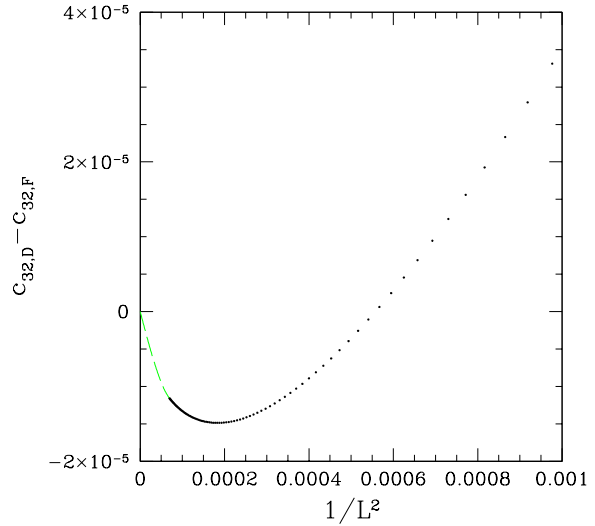}}
\caption{Coefficient $c_{32}$: difference between free and Dirichlet b.c.
}
\label{c32}
\end{figure}

The arguments of \cite{nnw} rely on two conjectures:

(i) equality of the thermodynamic limit of the PT coefficients, taken 
with free and Dirichlet b.c.:
\bee
\lim_{L\to\infty} c_{i,F}(L)=\lim_{L\to\infty} c_{i,D}(L)
\label{equal}
\ee
\\ 
(ii) a bound on the remainders $R_k$ in a region in the $(\beta,L)$
plane that can be considered as perturbative (for instance 
$L<\exp(\ln^2\beta)$. We have nothing to contribute to (ii), but 
some new computations confirming (i) beyond what was stated in \cite{nnw}.
All our computations, just as those of \cite{ps,nnw} refer to a special 
obervable, namely the nearest neighbor spin-spin scalar product located 
in the center of the lattice, i.e. $\cO=\vec s(0,0)\cdot \vec s(0,1)$. 

\begin{figure}[htb]
\centerline{\epsfxsize=8.0cm\epsfbox{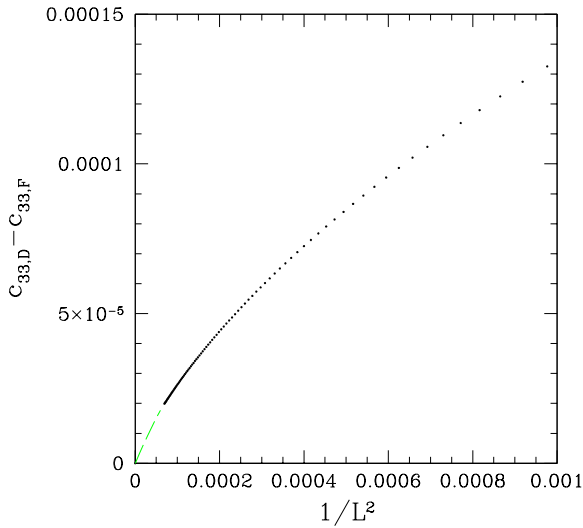}}
\caption{Coefficient $c_{33}$: difference between free and Dirichlet b.c.
}
\label{c33}
\end{figure}

In \cite{nnw} the authors state that for this observable they checked 
(\ref{equal}) for $i=2$. We confirmed this, and found that the difference  
$c_{i,D}(L)-c_{i,F}(L)$ goes to zero like $L^{-2}$.

The main new result we wish to report here is the computation of the 
third order coefficients $c_3$. For general $N$ we can 
write
\bee
c_{3,bc}=(N-1)c_{31,bc}+(N-1)^2c_{32,bc}+(N-1)^3c_{33,bc}\ .
\ee
We evaluated the coefficients $c_{31,bc}$, $c_{32,bc}$ and  $c_{33,bc}$
from $L=4$ up to $L=120$, where $bc$ stands with `$bc$' standing either 
for `$F$' (free) or `$D$' (Dirichlet). The differences $c_{3k,D}-c_{3k,F}$ 
($k=1,2,3$) are shown in the three figures below. In the figures the dots 
are the actual values, wheras the dashed lines are a spline through the 
points forced to hit the origin. The data strongly suggest that these 
differences go to zero like $O(L^{-2})$ for $L\to\infty$, even though the 
nonmonotonicity observed in $c_{32}$ is a little surprising.

In this brief report we cannot describe in detail how the PT coefficients
presented here were obtained; in fact the computations are quite complex, 
involving many lattice Feynman graphs that have to be evaluated with
complicated non-translation invariant propagators. The details 
of these computations, as well as a computation of the third order 
PT coefficients with s.i.b.c., confirming their logarithmic divergence
asserted by \cite{nnw} can be found in\cite{miguel}. 

\section{Discussion} 
By pushing the perturbative calculation of the nearest neighbor two-point 
function for the $O(N)$ model to order $\beta^{-3}$ we have provided some 
support for the arguments of \cite{nnw} in favor of the conventional view 
that the termwise thermodynamic limit of PT provides the correct 
asymptotic expansion in the thermodynamic limit. It should, however, be 
stressed that the main problem, namely control over the remainder terms 
$R_k$ remains open, and so does the important question of the status of PT 
in perturbatively asymptotically free theories.

The authors are grateful to P. Weisz for discussions. M.A. thanks T. Hahn 
for technical advice.

\end{document}